\documentclass[showpacs,twocolumn,superscriptaddress,prd,longbibliography,%
floatfix]{revtex4-1}

\usepackage{epsf}
\usepackage{amsmath}
\usepackage{amsfonts}
\usepackage{amssymb}
\usepackage{bm}
\usepackage{bbm}
\usepackage{nicefrac}
\usepackage{cancel}
\usepackage{xcolor}
\usepackage{maybemath}
\usepackage{pifont}
\usepackage{afterpage}
\usepackage{subfloat}

\usepackage{graphicx}
\usepackage{epsfig}
\usepackage{latexsym}
\usepackage{dcolumn}
\usepackage{pifont}

\definecolor{kellygreen}{rgb}{0.3, 0.73, 0.09}

\definecolor{garrosgreen}{rgb}{0.1, 0.4, 0.1}
\definecolor{dartmouthgreen}{rgb}{0.05, 0.5, 0.06}

\definecolor{duelferred}{rgb}{0.7, 0.2, 0.1}
\definecolor{cambridgeblue}{rgb}{0.1, 0.3, 1.0}
\definecolor{oxfordblue}{rgb}{0.05, 0.2, 0.7}

\def\dd{{\mathrm{d}}}

\newcommand{\calF}{\mathcal{F}}

\newcommand{\calM}{\mathcal{M}}

\newcommand{\calS}{\mathcal{S}}
\newcommand{\calT}{\mathcal{T}}

\newcommand{\calL}{\mathcal{L}}

\def\calL{{\mathcal L}}
\def\calF{{\mathcal F}}

\makeatletter

\newcommand{\Rmnum}[1]{\expandafter\@slowromancap\romannumeral #1@}
\makeatother

\newcolumntype{.}{D{x}{}{-1}}

\begin{document}

\title{Neutrino Pair Cerenkov Radiation for Tachyonic Neutrinos}


\author{Ulrich D. Jentschura}

\affiliation{Department of Physics,
Missouri University of Science and Technology,
Rolla, Missouri 65409, USA}

\affiliation{MTA--DE Particle Physics Research Group,
P.O.Box 51, H--4001 Debrecen, Hungary}

\author{Istv\'{a}n N\'{a}ndori}

\affiliation{MTA--DE Particle Physics Research Group,
P.O.Box 51, H--4001 Debrecen, Hungary}

\begin{abstract}
The emission of a charged light lepton
pair by a superluminal neutrino has
been identified as a major factor in the
energy loss of highly energetic neutrinos.
The observation of PeV neutrinos by IceCube
implies their stability against lepton pair 
Cerenkov radiation.
Under the assumption of a Lorentz-violating 
dispersion relation for highly energetic superluminal neutrinos,
one may thus constrain
the Lorentz-violating parameters. 
A kinematically different situation arises when 
one assumes a Lorentz-covariant, space-like 
dispersion relation for hypothetical tachyonic neutrinos, as an 
alternative to Lorentz-violating theories. 
We here discuss a hitherto neglected decay process,
where a highly energetic 
tachyonic neutrinos may emit other (space-like, tachyonic) neutrino pairs.
We find that the space-like dispersion relation implies the 
{\em absence} of a 
$q^2$ threshold for the production of a tachyonic 
neutrino-antineutrino pair,
thus leading to the dominant additional energy loss mechanism
for an oncoming tachyonic neutrino in the medium-energy domain. 
Surprisingly, the small absolute value of the decay rate and 
energy loss rate in the tachyonic model imply that 
these models, in contrast to the Lorentz-violating theories,
are not pressured by the cosmic PeV neutrinos registered
by the IceCube collaboration.  
\end{abstract}

\pacs{31.30.jh, 31.30.J-, 31.30.jf}

\maketitle


\newpage

%
%
\section{Introduction}
\label{sec1}

After early attempts at the construction of tachyonic neutrino 
theories~\cite{BiDeSu1962,DhSu1968,BiSu1969,Fe1967,Fe1978,ReMi1974,MaRe1980},
progress in the theoretical development 
was hindered by difficulties in the construction of a
viable field theory involving tachyons
(a particularly interesting argument was presented in Ref.~\cite{Bo1970}).
Despite the difficulties, work on tachyonic theories 
has continued up to this day, for both classical 
theories as well as spin-zero and spin-$1/2$ 
quantum theories~\cite{Ch2002,Re2009,Bi2009,Bo2009}.
A very interesting hypothesis was brought forward by
Chodos, Hauser and Kostelecky~\cite{ChHaKo1985}, 
who developed a tachyonic neutrino model based on the 
so-called tachyonic
Dirac equation. They recognized that 
a simple modification of the mass term in the 
Dirac equation, according to the replacement
$m \to \gamma^5 \, m$, induces a dispersion relation 
of the form $E = \sqrt{\vec p^{\,2} - m^2}$ 
(with the ``tachyonic'' sign in front of the mass term),
while preserving the spin-$1/2$ character of the equation.
Recently, it has been recognized~\cite{JeWu2012jpa}
that the modified Dirac Hamiltonian corresponding to the 
tachyonic solutions has a property known as 
pseudo--Hermiticity, which has been recognized as a 
viable generalization of the concept of Hermiticity,
for quantum mechanical systems~\cite{BeBo1998,%
BeDu1999,BeBoMe1999,BeWe2001,BeBrJo2002,
Mo2002i,Mo2002ii,Mo2002iii,Mo2003npb,%
JeSuZJ2009prl,JeSuZJ2010}.
Furthermore, the bispinor solution of the 
tachyonic equation have been determined~\cite{JeWu2013isrn},
and they have been shown to fulfill sum rules 
which enter the calculation of the 
time-ordered product of tachyonic field operators.
The tachyonic pseudo-Hermitian quantum dynamics
of wave packets composed of the bispinor 
solutions has been discussed in~\cite{JeWu2012jpa}.
A surprising feature of the tachyonic Dirac equation 
is the natural appearance of the fifth current 
in the equation. In particular, the appearance of 
$\gamma^5$ elevates the helicity 
basis to the most natural {\em ansatz} for the solution 
of the equation and induces parity-breaking in a natural way.
States with the ``wrong helicity'' are eliminated 
from the theory by a Gupta--Bleuler type 
condition~\cite{JeWu2013isrn}.

Just to fix ideas, we should point out here that 
the tachyonic neutrino differs from other 
faster-than-light neutrino models in that the 
dispersion relation is Lorentz-covariant.
Explicit breaking of the Lorentz symmetry 
may induce faster-than-light dynamics for neutrino 
wave packets, with a time-like four-vector product 
$p^\mu \, p_\mu > 0$ (see Refs.~\cite{KoLe2001,KoMe2012}). An example is the 
Lorentz-breaking dispersion relation $E = | \vec p | \, v$
with $v > 1$ (units with $\hbar = c = \epsilon_0$ are
used throughout this paper). This 
dispersion relation follows~\cite{CoGl2011,BeLe2012}
from a Lorentz violating ``metric''
$\tilde g_{\mu\nu} = {\rm diag}(1, -v, -v, -v)$.
A quite illuminating analysis of the model dependence
of the calculation~\cite{CoGl2011},
with reference to 
conceivably different forms of the interaction Lagrangian,
is given in Ref.~\cite{BeLe2012}.
By contrast, the tachyonic theory implies a 
space-like four-vector product
$p^\mu \, p_\mu = -m^2 < 0$, thus leaving
Lorentz symmetry intact and enabling the construction 
of bispinor solutions in the helicity basis~\cite{JeWu2013isrn}. 

Despite some ``seductive'' observations 
regarding the tachyonic neutrino model (most of all, pseudo--Hermiticity 
and natural emergence of the helicity eigenstates,
as well as the suppression of states with the 
``wrong'' helicity), 
any alternative neutrino model must also pass various 
other tests concerning the stability of highly energetic 
neutrinos against the emission of particle-antiparticle pairs.
The IceCube collaboration has registered ``big bird'', an  
$E_\nu =(2.004 \pm 0.236) \, {\rm PeV}$ 
highly energetic neutrino~\cite{AaEtAl2013,AaEtAl2014}. 
If neutrinos in this energy range are stable against 
lepton pair Cerenkov radiation, then this sets rather 
strict bounds on the values of 
the Lorentz-violating parameters~\cite{StSc2014,St2014}.
In a recent paper, lepton pair Cerenkov radiation 
has been analyzed as an
energy loss mechanism for high-energy tachyonic 
neutrinos~\cite{JeEh2016advhep}.
The kinematics in this case implies that 
the oncoming, decaying neutrino decays into a 
tachyonic state of lower energy, 
emitting an electron-positron pair [see Fig.~1(a)]. 
For the creation of an electron-positron pair, 
the threshold momentum for the virtual $Z^0$ 
boson is $q^2 = 4 m_e^2$ where $m_e$ is the 
electron mass.

For both the Lorentz-violating as well as the 
tachyonic neutrino models, one has not yet considered 
the additional decay and energy-loss channel which 
proceeds via a virtual $Z^0$ boson and has a neutrino-antineutrino 
pair (as opposed to an electron-positron pair) in the exit channel
[see Fig.~1(b)].
This process is not parametrically suppressed in 
comparison to the one with electrically charged 
particles in the exit channel, because of the 
weakly rather than electromagnetically 
interacting virtual particle (the $Z^0$ boson) 
in the middle.
For the Lorentz-violating theories, the 
kinematics in this case becomes involved 
because one has to implement Lorentz-violating 
parameters for all four particle in the process:
{\em (i)} the oncoming and exiting neutrino,
and {\em (ii)} the created neutrino-antineutrino
pair. Previous studies~\cite{CoGl2011,BeLe2012} 
have rather concentrated on the lepton pair Cerenkov radiation 
process as the dominant energy loss mechanism than the 
neutrino pair Cerenkov radiation; the kinematics 
in this case appears to be a lot
easier to analyze than for neutrino-pair
Cerenkov radiation.

\begin{figure} [t]
\begin{center}
\begin{minipage}{0.99\linewidth}
\begin{center}
\includegraphics[width=0.8\linewidth]{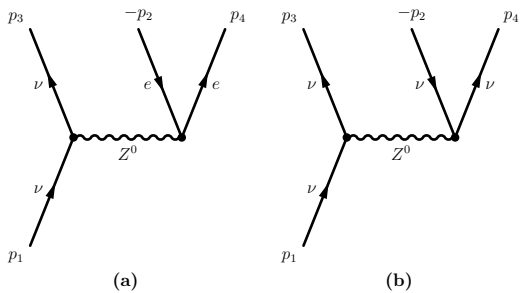}
\caption{\label{fig1} (Color online.)
The decay process in Fig.~(a) involves a faster-than-light 
neutrino decaying into an electron-positron pair,
with a threshold $q^2 = 4 m_e^2$ for the virtual $Z^0$ boson.
In Fig.~(b), a neutrino-antineutrino pair is emitted 
from an oncoming energetic neutrino.
Both processes are kinematically allowed if 
one assumes that neutrinos are tachyons described 
by a dispersion relation of the form
$p^\mu p_\mu = E^2 - \vec k^{\,2} = -m_\nu^2$.}
\end{center}
\end{minipage}
\end{center}
\end{figure}

For the tachyonic case, one needs to calculate the 
process of neutrino-antineutrino pair Cerenkov 
radiation in full tachyonic kinematics, for both 
the in and out states.
In particular, it is necessary to generalize the 
pair production threshold to the creation of a tachyonic
neutrino-antineutrino pair.
We organize this paper as follows.
In Sec.~\ref{sec2}, we derive the kinematic conditions 
for tachyon-antitachyon pair production.
The calculation of the threshold conditions and the
energy loss mechanism for neutrino pair Cerenkov radiation 
proceeds in Sec.~\ref{sec3}.
Consequences for tachyonic neutrino theories are 
summarized in Sec.~\ref{sec4}.

%
%
\section{Pair Production Threshold}
\label{sec2}

\begin{figure} [t]
\begin{center}
\begin{minipage}{0.99\linewidth}
\begin{center}
\includegraphics[width=0.8\linewidth]{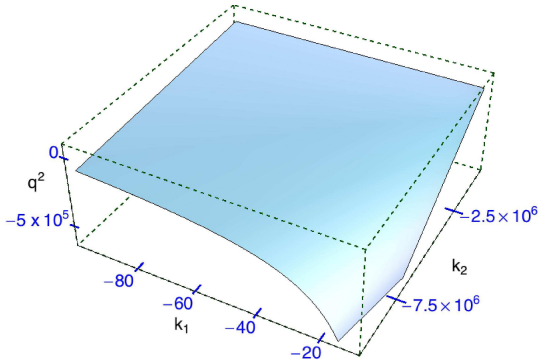}
\caption{\label{fig2} (Color online.)
Plot of $q^2$ given in Eq.~\eqref{q2plot}
for tachyonic pair production ($m_\nu = 1$) in the region 
$-100 < k_1 < 0$ and $-10^7 < k_2 < 0$,
demonstrating that $q^2$ may become
large and negative even for small tachyonic 
mass parameters, when one of the 
momenta is significantly larger than the other
[Eq.~\eqref{asymp}].}
\end{center}
\end{minipage}
\end{center}
\end{figure}

For two tardyonic (``normal'') particles of mass $m_e$, 
pair production threshold is reached when the pair is emitted 
collinearly, with two four-vectors 
$p^\mu = (E, \vec k) = (E_1, \vec k_1) = (E_2, \vec k_2)$
that fulfill
\begin{subequations}
\begin{align}
E =& \; \sqrt{ \vec k^{\,2} + m_e^2} \,,
\\[0.1133ex]
q^2 =& \; 4 p^\mu p_\mu = 4 \, (\vec k^2 + m_e^2) - 4 \, \vec k^{\,2} 
= 4\, m_e^2 \,.
\end{align}
\end{subequations}
The situation is completely different for the production of a tachyonic 
pair. Here, a well-defined lower threshold for $q^2$ is missing.
E.g., we have for the collinear pair with 
tachyonic mass parameter $E^2 - \vec k^{\,2} = -m_\nu^2$,
and $p^\mu = (E, \vec k) = (E_1, \vec k_1) = (E_2, \vec k_2)$,
\begin{subequations}
\begin{align}
E =& \; \sqrt{ \vec k^{\,2} - m_\mu^2} \,,
\\[0.1133ex]
q^2 =& \; 4 p^\mu p_\mu = 4 \, (\vec k^2 - m_\mu^2) - 4 \, \vec k^{\,2} 
= -4\, m_\mu^2 \,,
\end{align}
\end{subequations}
which is negative. For two neutrinos of different energy, 
emitted collinearly ($\vec k_1 = k_1 \, \hat{\rm e}_z$
and $\vec k_2 = k_2 \, \hat{\rm e}_z$), one has
\begin{subequations}
\begin{align}
E_1 =& \; \sqrt{ k_1^{\,2} - m_\mu^2} \,,
\qquad
E_2 = \sqrt{ k_2^{\,2} - m_\mu^2} \,,
\\[0.1133ex]
\label{q2plot}
q^2 =& \; 
\left( \sqrt{ k_1^{\,2} - m_\mu^2} +
\sqrt{ k_2^{\,2} - m_\mu^2} \right)^2 - 
( k_1 + k_2 )^2 \,.
\end{align}
\end{subequations}
In the limit of a small tachyonic mass parameter $m_\nu$,
a Taylor expansion of the latter term leads to the expression
\begin{align}
\label{asymp}
q^2 = -\left( 2 + \frac{k_1}{k_2} + \frac{k_2}{k_1} 
\right) \, m_\nu^2 + {\cal O}(m_\nu^4) \,.
\end{align}
In the limits $k_1 \to 0$, $k_2 \to \infty$ or alternatively
$k_1 \to \infty$, $k_2 \to 0$,
the latter expression may assume very large negative numerical values
(see also Fig.~\ref{fig2}). 
There is thus no lower threshold for tachyonic pair production,
expressed in $q^2$.

One might ask if arbitrarily large $q^2$ are compatible 
with the relativistic tachyonic pair production kinematics.
In order to answer this question,
we consider the production of an anti-collinear pair,
\begin{subequations}
\begin{align}
\vec k_1 =& \; k_1 \, \hat{\rm e}_z \,,
\qquad
\vec k_2 = -k_2 \, \hat{\rm e}_z \,,
\\[0.1133ex]
E_1 =& \; \sqrt{ k_1^{\,2} - m_\mu^2 } \,,
\qquad
E_2 = \sqrt{ k_2^{\,2} - m_\mu^2 } \,,
\\[0.1133ex]
q^2 =& \; 
\left( \sqrt{ k_1^{\,2} - m_\mu^2 } +
\sqrt{ k_2^{\,2} - m_\mu^2} \right)^2 - 
( k_1 - k_2 )^2 
\\[0.1133ex]
=& \; 4 k_1 k_2 + {\cal O}(m_\nu^2) \,.
\end{align}
\end{subequations}
For large $k_1$ and $k_2$, this expression assumes arbitrarily large
positive numerical values.

The conclusion is that the tachyonic kinematics 
do not exclude any range of $q^2$ from the kinematically 
allowed range of permissible momentum transfers, 
and neutrino pair Cerenkov radiation (or, more generally, 
tachyon--antitachyon pair production) is 
allowed in the entire range 
\begin{equation}
\label{q2range}
-\infty < q^2 < \infty \,,
\qquad
q^0 > 0 \,.
\end{equation}
The latter condition only ensures that the 
energy emitted into the pair is positive.
For a decaying tachyonic neutrino, 
the condition~\eqref{q2range} implies that 
there is no lower energy threshold for 
the production of a tachyon-antitachyon pair 
from an oncoming neutrino, within the 
process depicted in Fig.~\ref{fig1}(b).


%
%
\section{Calculation of the Pair Production}
\label{sec3}

We calculate the decay width of the incoming tachyonic neutrino,
in the lab frame, employing a relativistically covariant
(tachyonic) dispersion relation,
with both incoming as well as outgoing neutrinos
on the tachyonic mass shell
($E_i = \sqrt{\vec k_i^2 - m_\nu^2}$ for $i=1,2,3,4$),
in the conventions
of Fig.~\ref{fig1}. In the lab frame, the decay rate is
\begin{align}
\label{GammaTacStart}
\Gamma =& \;  \frac{1}{2 E_1} \,
\int \frac{\dd^3 p_3}{(2 \pi)^3 \, 2 E_3}
\, \left(
\int \frac{\dd^3 p_2}{(2 \pi)^3 \, 2 E_2}
\int \frac{\dd^3 p_4}{(2 \pi)^3 \, 2 E_4}
\right.
\nonumber\\
& \; \left. \times (2 \pi)^4 \, \delta^{(4)}( p_1 - p_3 - p_2 - p_4 ) \,
\left[ {\widetilde \sum}_{\rm spins} | \calM |^2 \right] 
\right) \,.
\end{align}
Here, ${\widetilde \sum}_{\rm spins}$ refers to the
specific way in which the average over the
oncoming helicity states, and the outgoing helicities,
needs to be carried out for tachyons~\cite{JeEh2016advhep}.

We use the Lagrangian 
\begin{equation}
\calL = - \frac{g_w}{4 \, \cos\theta_W} \, 
\left[ \overline \nu \, 
\gamma^\mu (1 - \gamma^5) \, \nu \right] \, Z_\mu  \,,
\end{equation}
where $\theta_W$ is the Weinberg angle, 
$Z_\mu$ is the $Z^0$ boson field,
and $\nu$ is the neutrino field.
The effective four-fermion interaction is 
\begin{equation}
\calL = \frac{G_F}{2 \sqrt{2}} \,
\left[ \overline \nu \, 
\gamma^\mu (1 - \gamma^5) \, \nu \right] \,
\left[ \overline \nu \, 
\gamma^\mu (1 - \gamma^5) \, \nu \right] \,,
\end{equation}
where $G_F$ is the Fermi coupling constant.
The matrix element $\calM$ is 
\begin{align}
\calM =& \; \frac{G_F}{2 \sqrt{2}} \,
\left[ \overline u^\calT(p_3) \, \gamma_\lambda \, ( 1- \gamma^5) \,
u^\calT(p_1) \right] \,
\nonumber\\[0.1133ex]
& \; \times 
\left[ \overline u^\calT(p_4) \, \gamma^\lambda \, 
( 1 - \gamma^5 ) \, v^\calT(p_2) \right] \,,
\end{align}
where $u^\calT(p)$ is a tachyonic positive-energy 
bispinor (particle) solution, while 
$v^\calT(p)$ is a tachyonic negative-energy 
(antiparticle) solution. The positive-energy solutions 
read as follows~\cite{JeWu2013isrn},
\begin{subequations}
\begin{align}
u^\calT_+(\vec k) = & \;
\left( \begin{array}{c} \sqrt{ | \vec k | + m} \; a_+(\vec k) \\
\sqrt{ | \vec k | - m} \; a_+(\vec k)
\end{array} \right)  \,,
\\[0.1133ex]
u^\calT_-(\vec k) = & \;
\left( \begin{array}{c} \sqrt{ | \vec k | - m} \; a_-(\vec k) \\
- \sqrt{ | \vec k | + m} \; a_-(\vec k)
\end{array} \right)  \,,
\end{align}
while the negative-energy solutions are given by 
\begin{align}
v^\calT_+(\vec k) = & \;
\left( \begin{array}{c} -\sqrt{ | \vec k | - m} \; a_+(\vec k) \\
-\sqrt{ | \vec k | + m} \; a_+(\vec k)
\end{array} \right)  \,,
\\[0.1133ex]
v^\calT_-(\vec k) = & \;
\left( \begin{array}{c} - \sqrt{ | \vec k | + m} \; a_-(\vec k) \\
\sqrt{ | \vec k | - m} \; a_-(\vec k)
\end{array} \right)  \,,
\end{align}
\end{subequations}
where we identify the on-shell spinors 
$u^\calT(p)$ with the 
$u^\calT(\vec k)$, where $p^\mu = (E, \vec k)$ and 
$E = \sqrt{\vec k^{\,2} - m_\nu^2}$.
The symbols $a_\pm(\vec k)$ denote the fundamental
helicity spinors (see p.~87 of Ref.~\cite{ItZu1980}).
We note that the helicity of the antineutrino solution 
$v^\calT_-(\vec k)$ is positive, 
while in the massless limit, 
it has negative chirality.


For the tachyonic spin sums, one has the
following sum rule for the positive-energy spinors~\cite{JeWu2012epjc},
\begin{multline}
\label{sumrule1}
\sum_\sigma (-\sigma )\;
u^\calT_\sigma(\vec k) \otimes \overline u^\calT_\sigma(\hat k) \, \gamma^5 
\\ 
= \sum_\sigma \left(-\vec\Sigma \cdot \hat k \right)\;
u^\calT_\sigma(\vec k) \otimes \overline u^\calT(\vec k) \, \gamma^5 =
\cancel{p} - \gamma^5 \, m \,,
\end{multline}
where $\hat k = \vec k / | \vec k|$ is the unit vector
in the $\vec k$ direction.
Upon promotion to a four-vector, one has 
$\hat k^\mu = (0, \hat k)$.
The sum rule can thus be reformulated as
\begin{multline}
\label{sumrule2}
\sum_\sigma u^\calT_\sigma(p) \otimes \overline u^\calT_\sigma(p)
= \left( -\vec\Sigma \cdot \hat k \right) \,
( \cancel{p} - \gamma^5 \, m_\nu) \; \gamma^5 \\
= - \gamma^5 \, \gamma^0 \, \gamma^i \, {\hat k}^i \;
( \cancel{p} - \gamma^5 \, m_\nu) \; \gamma^5 \\
= -\cancel{\tau} \, \gamma^5 \, \cancel{\hat{k}} \;
( \cancel{p} - \gamma^5 \, m_\nu) \; \gamma^5 \,,
\end{multline}
where $\tau = (1,0,0,0)$ is a time-like unit vector.

In Refs.~\cite{JeWu2012epjc,JeWu2013isrn}, it has been
argued that a consistent formulation of the
tachyonic propagator is achieved when we postulate that the
right-handed neutrino states, and the left-handed
antineutrino states, acquire a negative Fock-space norm
after quantization of the tachyonic spin-$1/2$ field.
Hence, in order to calculate the decay process of
an oncoming, left-handed, positive-energy neutrino,
we must first project onto negative-helicity
states, according to Ref.~\cite{JeEh2016advhep},
\begin{multline}
\label{sumrule3}
\frac12 \, \left( 1 - \vec\Sigma \cdot \hat k \right) \;
\sum_\sigma u^\calT_\sigma(\vec k) \otimes 
\overline u^\calT_\sigma(\vec k) 
\\
= u_{\sigma=-1}(p) \otimes \overline u_{\sigma=-1}(p) 
\\
= \frac12 \, \left( 1 - \cancel{\tau} \, \gamma^5 \, 
\cancel{\hat{k}} \right) \,
( \cancel{p} - \gamma^5 \, m_\nu) \; \gamma^5 \,.
\end{multline}
The squared and spin-summed matrix element 
for the tachyonic decay process thus is
\begin{align}
\label{tacTrace}
\mathop{{\widetilde \sum}}_{\rm spins} | \calM |^2 =& \;
\frac{G_F^2}{8} \, T_{13} \, T_{24} 
= \frac{G_F^2}{8} \, \calS(p_1, p_2, p_3, p_4) \,,
\end{align}
where the latter identity provides for an 
implicit definition of the function 
$\calS(p_1, p_2, p_3, p_4)$.
The traces $T_{13}$ and $T_{24}$ are
\begin{subequations}
\begin{align}
T_{13} =& \; {\rm Tr} \left[ 
\frac12 \, \left( 1 - \cancel{\tau} \, \gamma^5 \cancel{\hat{k}}_3 \right) \;
(\cancel{p}_3 - \gamma^5 \, m_{\nu}) \, \gamma^5 \, \gamma_\lambda \,
( 1- \gamma^5) \right.
\nonumber\\[0.1133ex]
& \; \left. \times \frac12 \, 
\left( 1 - \cancel{\tau} \, \gamma^5 \, \cancel{\hat{k}}_1 \right) \;
(\cancel{p}_1  - \gamma^5 \, m_{\nu} ) \, \gamma^5 \, \gamma_\nu  \,
( 1- \gamma^5) \right] \,,
\\[0.1133ex]
T_{24} =& \; 
{\rm Tr} \left[ 
\frac12 \, 
\left( 1 - \cancel{\tau} \, \gamma^5 \cancel{\hat{k}}_4 \right) \,
( \cancel{p}_4 - \gamma^5 \, m_\nu ) \, \gamma^5 \,
\gamma^\lambda ( 1 - \gamma^5 ) \right.
\nonumber\\[0.1133ex]
& \; \left. \times \frac12 \, 
\left( 1 - \cancel{\tau} \, \gamma^5 \cancel{\hat{k}}_2 \right) \,
( \cancel{p}_2 + \gamma^5 \, m_\nu ) \, \gamma^5 \,
\gamma_\lambda ( 1 - \gamma^5 ) 
\right] \,.
\end{align}
\end{subequations}
We have chosen the convention to denote the 
by $p_2$ the momentum of the outgoing antiparticle.

For the outgoing pair, we use the fact that the 
helicity projector is approximately equal to the 
chirality projector in the high-energy limit, 
which simplifies the Dirac gamma trace somewhat.
On the tachyonic mass shell, one has
$p_1^2 = p_2^2 = p_3^2 = p_4^2 = -m_\nu^2$.
After the trace over the Dirac $\gamma$ matrices,
some resultant scalar products vanish, e.g.,
the scalar product of the time-like unit
vector $\tau$ and the space-like unit vector
($\tau \cdot \hat k = 0$).

The result of the Dirac $\gamma$ traces from Eq.~\eqref{tacTrace}
is inserted into Eq.~\eqref{GammaTacStart},
and the $\dd^3 p_2$ and $\dd^3 p_4$ integrals
are carried out using the following formulas,
\allowdisplaybreaks
\begin{subequations}
\begin{align}
I(q) =& \;
\int \frac{\dd^3 p_2}{2 E_2}
\int \frac{\dd^3 p_4}{ 2 E_4}
\delta^{(4)}( q - p_2 - p_4 ) 
\nonumber\\[0.1133ex]
=& \; \frac{\pi}{2} \, \sqrt{1 + \frac{4 \, m_\nu^2}{ q^2 }} \,,
\\[0.133ex]
J_{\lambda\rho}(q) =& \;
\int \frac{\dd^3 p_2}{2 E_2}
\int \frac{\dd^3 p_4}{ 2 E_4}
\delta^{(4)}( q - p_2 - p_4 ) \,
\left( p_{2 \lambda} \; p_{4 \rho} \right)
\nonumber\\[0.1133ex]
=& \; \sqrt{1 + \frac{4 \, m_\nu^2}{q^2}} \,
\left[ g_{\lambda \rho} \, \frac{\pi}{24} \,
\left( q^2 + 4 m_\nu^2 \right) 
\right.
\nonumber\\[0.1133ex]
& \; \left. + q_\lambda \, q_\rho \,
\frac{\pi}{12} \,
\left( 1 - \frac{2 m_\nu^2}{q^2} \right) \right] \,,
\\[0.133ex]
K(q) =& \;
\int \frac{\dd^3 p_2}{2 E_2}
\int \frac{\dd^3 p_4}{ 2 E_4}
\delta^{(4)}( q - p_2 - p_4 ) \,
\left( p_2 \cdot p_4 \right)
\nonumber\\[0.1133ex]
=& \; \frac{\pi}{4} \,
\sqrt{1 + \frac{4 \, m_\nu^2}{q^2}} \,
\left( q^2 + 2 m_\nu^2 \right) \,.
\end{align}
\end{subequations}
After the $\dd^3 p_2$ and $\dd^3 p_4$ integrations,
we are left with an expression of the form
\begin{align}
\label{GammaF}
\Gamma =& \; \frac{G_F^2}{8} \frac{1}{(2 \pi)^5} \,
\int\limits_{q^2 > 4 m_e^2} \frac{\dd^3 p_3}{2 E_3} \,
\calF(p_1, p_3) \,,
\end{align}
where
\begin{align}
\label{defcalF}
\calF(p_1, p_3) =
\int \frac{\dd^3 p_2}{2 E_2}
\int \frac{\dd^3 p_4}{ 2 E_4}
\delta^{(4)}( p_1 - p_2 - p_3 - p_4 ) 
\nonumber\\[0.1133ex]
\times \calS(p_1, p_2, p_3, p_4) \,.
\end{align}
Both the expressions for $\calS(p_1, p_2, p_3, p_4)$
as well as $\calF(p_1, p_3)$ are too lengthy to be displayed
in the context of the current paper. 

For the kinematics, we assume that
\begin{align}
p_1^\mu =& \; (E_1, 0,0, k_1) \,,
\nonumber\\[0.1133ex]
p_3^\mu =& \; (E_3, 
k_3 \, \sin\theta \, \cos\varphi , 
k_3 \, \sin\theta \, \sin\varphi , 
k_3 \, \cos\theta) \,,
\nonumber\\[0.1133ex]
E_3^2 - k_3^2 =& \; -m_\nu^2 \,,
\qquad
k_3 > m_\nu \,.
\end{align}
The condition $k_3 > m_\nu$ is naturally imposed for
tachyonic kinematics.
The squared four-momentum transfer then reads as
\begin{align}
q^2 =& \; 2 \, \left( 
\sqrt{E_1^2 + m_\nu^2} \sqrt{E_3^2 + m_\nu^2} \, \cos \theta 
-E_1 E_3 - m_\nu^2 \right) 
\nonumber\\[0.1133ex]
=& \; 2 \, \left(
k_1 \, k_3 \, u
-\sqrt{k_1^2 - m_\nu^2} \, \sqrt{k_3^2 - m_\nu^2} - m_\nu^2 \right) \,,
\end{align}
where it is convenient to define $u = \cos\theta$.

The integrations are done
with the kinematic conditions that all $0 < E_3 < E_1$, 
and all $q^2 = (p_2 + p_4)^2$ for the pair are allowed
(see Sec.~\ref{sec2}), leading to 
\begin{align}
\Gamma =& \; \frac{G_F^2}{8} \frac{1}{(2 \pi)^5} \,
\int\limits_0^{2 \pi} \dd \varphi \,
\int\limits_{k_3 = m_\nu}^{k_{\rm max}}
\frac{\dd k_3 \, k_3^2}{2 E_3} \,
\int\limits_{-1}^1 \dd u \, \calF(E_1, E_3, u) 
\nonumber\\[0.1133ex]
=& \; \frac{G_F^2}{16} \frac{1}{(2 \pi)^4} \,
\int\limits_{0}^{E_1}
\dd E_3 \, \sqrt{E_3^2 + m_\nu^2}
\int\limits_{-1}^1 \dd u \, \calF(E_1, E_3, u) \,,
\end{align}
where $k_{\rm max} = \sqrt{E_1^2 + m_\nu^2}$
and we have used the identity
\begin{equation}
\dd k_3 \, k_3 = \dd E_3 \, E_3 \,,
\qquad
k_3 = \sqrt{E_3^2 + m_\nu^2} \,.
\end{equation}
The differential energy loss,
for a particle traveling at velocity $v_\nu \approx c$,
undergoing a decay with energy loss $E_1 - E_3$,
due to the energy-resolved decay rate
$(\dd \Gamma/\dd E ) \, \dd E$, in time $\dd t = \dd x/c$,
reads as follows,
\begin{equation}
\dd^2 E_1 = -(E_1 - E_3) \, \frac{\dd \Gamma}{\dd E_3}
\dd E_3 \, \frac{\dd x}{c} \,.
\end{equation}
Now we set $c = 1$, divide both sides of the equation
by $\dd x$ and integrate over the energy $E_3$ 
of the outgoing particle. One obtains
\begin{equation}
\frac{\dd E_1}{\dd x} =
- \int \dd E_3 \, (E_1 - E_3) \, \frac{\dd \Gamma}{\dd E_3} \,.
\end{equation}
Hence, the energy loss rate is obtained as
\begin{align}
\frac{\dd E}{\dd x} =& \;
 -\frac{G_F^2}{4} \frac{1}{(2 \pi)^4} \,
\int\limits_{0}^{E_1}
\dd E_3 \sqrt{E_3^2 + m_\nu^2} \, (E_1 - E_3) 
\nonumber\\[0.1133ex]
& \; \times \int\limits_{-1}^1 \dd u \, \calF(E_1, E_3, u) \,.
\end{align}
After a long, and somewhat tedious integration one finds the
following expressions,
\begin{subequations}
\begin{align}
\label{Gamma}
\Gamma =& \; 
\dfrac13 \; \dfrac{G_F^2 \, m_\nu^4}{192 \pi^3} \; E_1 \,,
\\[4ex]
\label{dEdx}
\frac{\dd E_1}{\dd x} =& \; 
\dfrac13 \; \dfrac{G_F^2 \, m_\nu^4}{192 \pi^3} \; E^2_1 \,.
\end{align}
\end{subequations}
These formulas are valid for $E_1 \gg m_\nu$, which 
is easily fulfilled for all neutrino masses $m_\nu$.
There is no threshold energy; i.e., formulas~\eqref{Gamma}
and~\eqref{dEdx} are, in particular, valid in the range 
$E_1 \gtrsim 1 \, {\rm eV} \gg m_\nu$.
Parametrically, they are of the same
order-of-magnitude as those given in Ref.~\cite{JeEh2016advhep}
for (charged) lepton pair Cerenkov radiation, 
but the threshold is zero for the neutrino pair emission.
Hence, neutrino pair emission is the dominant decay channel
in the medium-energy domain, for an oncoming tachyonic neutrino 
flavor eigenstate.

%
%
\section{Discussion and Conclusions}
\label{sec4}

In principle, tachyonic spin-$1/2$ theories 
have a number of properties which make them 
more attractive than their spin-zero counterparts.
One distinctive feature is that the mass parameters
enters only linearly in the 
Lagrangian~\cite{JeWu2013isrn}, thus preventing the vacuum 
from becoming manifestly unstable against tachyon-antitachyon
pair production.
Also, it has been possible to calculate the 
time-ordered product of field operators,
which leads to the Feynman propagator
of the tachyonic field~\cite{JeWu2012epjc,JeWu2013isrn}.
One also observes that the generalized Dirac 
Hamiltonian for the tachyonic spin-$1/2$ fields 
is pseudo-Hermitian, so that it becomes 
possible to formulate the quantum dynamics
of tachyonic wave packets without having to 
overcome unsurmountable challenges~\cite{JeWu2012jpa}.
In Ref.~\cite{JeEtAl2014}, 
it has been argued that in view of the 
small neutrino interaction cross sections,
it would be difficult to transport information 
faster than the speed of light using a neutrino
beam, if neutrino are just a bit superluminal
(tachyonic). 
The sign of the mass square of neutrinos has not 
yet been determined experimentally, in contrast to
differences of mass squares among neutrino 
flavor eigenstates.

Here, we calculate the decay rate and energy loss rate, 
for a hypothetically tachyonic neutrino flavor, against neutrino-pair 
Cerenkov radiation. It needs to be checked if the 
absence of a threshold would lead to 
a disagreement with high-energy data on 
neutrinos of cosmic origin.
In fact, the IceCube
experiment~\cite{AaEtAl2013,AaEtAl2014} has observed 37 neutrinos having
energies $E_\nu>10 \, {\rm TeV}$ during the 
first three years of data taking. Three of these events
(``Ernie'', ``Bert'' and ``Big Bird'')
had energies $E_\nu > 1 \, {\rm PeV}$, while ``Big Bird''
is famous for having an energy of 
$E_\nu= (2.004 \pm 0.236)\,{\rm PeV}$.
A blazar has been identified as a 
possible source of this highly energetic neutrino~\cite{KaEtAl2016}.
Neutrinos registered by IceCube have to
``survive'' the possibility of energy loss by decay,
and if they are tachyonic, then lepton and neutrino 
pair Cerenkov radiation processes become kinematically 
allowed. 

The results given in Eqs.~\eqref{Gamma} 
and~\eqref{dEdx} for the 
decay rate and energy loss rate due to 
neutrino pair Cerenkov radiation are not subject to a threshold energy;
parametrically are of the same order-of-magnitude 
as those given for lepton pair Cerenkov radiation in 
Ref.~\cite{JeEh2016advhep}, but the threshold energy is zero.
Let us estimate the relative energy loss due to 
neutrino pair Cerenkov radiation
over a distance
\begin{equation}
L = 15 \times 10^9 \, {\rm ly} = 1.42 \times 10^{26} \, {\rm m}  \,,
\end{equation}
assuming a (relative large) neutrino mass
parameter of $m_0 = 10^{-2} \, {\rm eV}$.
One obtains for the relative energy loss
according to Eq.~\eqref{dEdx},
\begin{equation}
\frac{L}{E_1} \, \frac{\dd E_1}{\dd x} = 
\dfrac13 \; \dfrac{G_F^2 \, m_0^4}{192 \pi^3} \, E_1 \, L =
5.02 \times 10^{-20} \, \frac{E_1}{{\rm MeV}} \,.
\end{equation}
This means that even at the large ``Big Bird'' 
energy of $E_\nu= (2.004 \pm 0.236)\,{\rm PeV}$,
the relative energy loss over 15 billion 
light years does not exceed $5$ parts in $10^{11}$,
which is negligible.

The decay rate is obtained as follows
(again, assuming that $m_\nu = 10^{-2} \, {\rm eV}$),
\begin{equation}
\Gamma =
\dfrac13 \; \dfrac{G_F^2 \, m_0^4}{192 \pi^3} \; E_1 
= 1.06 \times 10^{-37} \, \left( \frac{E_1}{{\rm MeV}} \right) \, 
\left( \frac{\rm rad}{\rm s} \right) \,.
\end{equation}
Even for ``Big Bird'', this means that the 
decay rate does not exceed
$2.12 \times 10^{-28} \, \frac{\rm rad}{\rm s}$,
which is equivalent to a lifetime of 
$\sim 10^{20}$~years, far exceeding the age
of the Universe. The neutrino pair Cerenkov radiation process,
even if threshold-less, 
has such a low probability due to the 
weak-interaction physics involved,
that it cannot constrain the tachyonic 
models. Indeed, even for relatively large
tachyonic neutrino mass parameters of the 
order of $10^{-2} \, {\rm eV}$, and for the 
largest  neutrino energies observed, 
the process leads only to a vanishingly 
small relative energy loss for an 
oncoming neutrino of cosmic origin 15\,billion\,light\,years
away. The lifetime of the tachyonic neutrino 
far exceeds the age of the Universe.
Our quick estimate shows that ``Big Bird'' 
would have survived the travel from the 
blazar PKS B1424-418~(see Ref.~\cite{KaEtAl2016}).
In other words, neutrino pair Cerenkov 
radiation does not pressure the
tachyonic neutrino hypothesis.

We thus take the opportunity here to 
correct claims recently made by 
by one of us (U.D.J.) in Ref.~\cite{JeEh2016advhep},
where a hypothetical cutoff of cosmic neutrino 
spectrum at the Big Bird energy was 
related to the threshold energy for (charged) lepton pair Cerenkov 
radiation, and thus, to a neutrino mass parameter.
In Ref.~\cite{JeEh2016advhep}, it was 
overlooked that {\em (i)} a further decay process 
exists for tachyonic neutrinos which is not 
subject to a threshold condition,
and {\em (ii)} that the absolute value of 
{\em both} (charged) lepton as well as 
neutrino pair Cerenkov radiation is too small
(both above as well as below threshold)
to lead to any appreciable energy loss 
of an oncoming tachyonic neutrino flavor eigenstate,
over cosmic distances and time scales.
Hence, it is not possible, 
in contrast to the conclusions of Ref.~\cite{JeEh2016advhep},
to relate the lepton pair threshold to 
the tachyonic mass parameter.
The (more optimistic) conclusion thus 
is that neither lepton nor 
neutrino pair Cerenkov radiation processes 
pressure the tachyonic model.

However, for the Lorentz-violating models,
important limits on the available parameter
space have been set in Refs.~\cite{StSc2014,St2014},
based on (charged) lepton pair Cerenkov radiation alone.
Roughly speaking, the reason for the pressure on 
the Lorentz-violating models is that even 
small Lorentz violations at PeV energies correspond
to high ``virtualities'' of the superluminal particles
and hence, relatively large (energy-dependent) mass parameters.
It is quite imperative that 
the additional decay process studied here should also be 
calculated for the different kinematic conditions
in Lorentz-violating models, where it will further
limit the available parameter space for the 
Lorentz-violating parameters.
Note that, e.g., 
employing a Lorentz-violating dispersion relation 
$E = v \, |\vec p|$~\cite{CoGl2011,BeLe2012},
with $v > 1$, a quick calculation shows the absence
of a neutrino pair production threshold in the 
Lorentz-violating model; the reason being simple: 
namely, one has $E \to 0$ for $|\vec p| \to 0$, 
and it thus becomes possible to generate 
Lorentz-violating neutrino pairs with near-zero 
four-momenta. The additional decay process 
uncovered here thus has the potential 
of fundamentally changing the 
bounds to be inferred for the Lorentz-violating 
parameters, from the cosmic high-energy neutrinos,
within the Lorentz-violating models.

To conclude,
the Lorentz-violating model is pressured at high energies,
where even numerically tiny values of the Lorentz-violating
parameters induce large deviations from the light-like
dispersion relation, corresponding to a numerically
large value of the ``effective mass'' $m_*$ with
$E^2 - \vec p^{\,2} = m_*^2 = (v^2 - 1) \, \vec p^{\,2}$
(where we assume the dispersion relation $E = v \, |\vec p|$
given in Refs.~\cite{CoGl2011,BeLe2012}). By contrast, the
tachyonic model is fully compatible with astrophysical data
collected at high energies, while the tachyonic dispersion
relation predicts noticeable deviations from the
speed of light only for comparatively low-energy neutrinos.
A proposal to test the tachyonic hypothesis, in the
low-energy domain, has recently been published in Ref.~\cite{JeEtAl2014}.
Finally, we also refer to Ref.~\cite{JeEtAl2014} for clarifying remarks on
general aspects of the tachyonic model.

\acknowledgments{}

The authors acknowledge helpful conversations with 
R.~Ehrlich. This research has been supported
by the NSF (grant PHY--1403937) and
by a J\'anos Bolyai Research Scholarship
of the Hungarian Academy of Sciences.

{\em The authors declare that there is no conflict of interest regarding
the publication of this paper.”}

\end{document}